\documentclass[twocolumn,showpacs,preprintnumbers,amsmath,amssymb]{revtex4}
\usepackage{graphicx}
\usepackage{dcolumn}
\usepackage{color}
\usepackage{bm}
\usepackage{amsmath,amssymb,graphicx}
\usepackage{epsfig}
\usepackage{enumerate}
\usepackage{array}
\usepackage{multirow}
\newtheorem{theorem}{Theorem}

\newtheorem{prop}{Proposition}
\newtheorem{corollary}{Corollary}

\DeclareUnicodeCharacter{2212}{-}

\begin{document}

\title{Contact geometry and quantum thermodynamics of nanoscale steady states}
\author{Aritra Ghosh\footnote{E-mail: ag34@iitbbs.ac.in}, Malay Bandyopadhyay\footnote{E-mail: malay@iitbbs.ac.in} and Chandrasekhar Bhamidipati\footnote{E-mail: chandrasekhar@iitbbs.ac.in}}

\affiliation{School of Basic Sciences, Indian Institute of Technology Bhubaneswar,\\  Jatni, Khurda, Odisha, 752050, India}

\vskip-2.8cm
\date{\today}
\vskip-0.9cm

\begin{abstract}
We develop a geometric formalism suited for describing the quantum thermodynamics of a certain class of nanoscale systems (whose density matrix is expressible in the McLennan--Zubarev form) at any arbitrary non-equilibrium steady state. It is shown that the non-equilibrium steady states are points on control parameter spaces which are in a sense generated by the steady state Massieu--Planck function. By suitably altering the system's boundary conditions, it is possible to take the system from one steady state to another. We provide a contact Hamiltonian description of such transformations and show that moving along the geodesics of the friction tensor results in a minimum increase of the free entropy along the transformation. The control parameter space is shown to be equipped with a natural Riemannian metric that is compatible with the contact structure of the quantum thermodynamic phase space which when expressed in a local coordinate chart, coincides with the Schl\"{o}gl metric. Finally, we show that this metric is conformally related to other thermodynamic Hessian metrics which might be written on control parameter spaces. This provides various alternate ways of computing the Schl\"{o}gl metric which is known to be equivalent to the Fisher information matrix.
\end{abstract}

\pacs{73.23.−b, 72.10.−d, 73.63.Kv}

\maketitle

\section{Introduction}
\label{sec:Intro}
Boltzmann and Gibbs formulated the prescription of equilibrium statistical mechanics by providing the appropriate density matrix (or density operator) of a system kept at temperature \(T\) which is given by the canonical distribution,
\begin{equation}
\rho = \frac{e^{-\beta H}}{Z}
\end{equation}
where $\beta =T^{-1}$ (the Boltzmann constant \(k_B\) is set to unity throughout the paper) and \(Z\) is the normalizing partition function \cite{gibbs}. For systems in (weak) contact with both a thermostat and a particle reservoir, the grand canonical distribution describes the statistical state of the system. Hence, this canonical or grand canonical prescription is the foundation stone of equilibrium statistical physics and these phase space distributions govern the laws of classical equilibrium thermodynamics. Thus, equilibrium statistical mechanics is all about finding the correct density matrix or the phase space distribution of a system enabling us to compute various equilibrium properties. On the other hand, we can utilize the linear response theory \cite{kubo1,kubo2} and Onsager reciprocity relationships \cite{onsager1,onsager2} to study systems close to thermodynamic equilibrium. But, the majority of biological, technological and even cosmological systems fall into the category of systems far from equilibrium where there is no general formalism to obtain the correct density matrix of the system. This active and vibrant research area is primarily centered around transport phenomena \cite{Bird, Glasgow}, chemical transformations \cite{Rao} and autonomous machines \cite{Benenti,Guarnieri} (both classical and quantum  perspective). Although thermodynamic relations away from equilibrium in general situations have been explored before (see for example~\cite{Oono98,Sasa06,Komatsu08,Saito11c,Hyldgaard12}) including situations involving non-interacting transport~\cite{Ludovico14,Esposito15,Topp15,Bruch16,Stafford17}, this still leaves a lot to be formulated, as definitions of basic thermodynamic quantities still remain to be understood. Some important results in truly non-equilibrium steady state situations for certain class of models have recently been revealed in~\cite{taniguchi}, which will form the basis for our analysis and to which we return in section-(\ref{section2}).

\smallskip

 Quite remarkably, it has been found in the recent years that thermodynamic properties are consistent with the quantum properties of nanoscale systems which are far from the thermodynamic limit and may even contain a single particle. The advancement in technologies and experimental techniques have enabled us to measure the thermodynamic properties of microscopic systems \cite{a1,b1,c1,d1}. For example Collin et al. \cite{collin} measured the work performed on a single RNA hairpin using optical tweezers and also found the equilibrium free energy change using the work fluctuation relations from such out of equilibrium measurements. Moreover, the Jarzynski equality \cite{jar1,jar2,h,i} relating different thermodynamic quantities for systems driven far from equilibrium and fluctuation theorems related to entropy production probability during a finite period of time \cite{d,e,f,g} have been verified theoretically as well as experimentally. Interestingly, some of these ideas of non-equilibrium physics find applications in machine learning \cite{machine} as well as learning and inference problems \cite{h1,i1}. A more profound problem in modern non-equilibrium thermodynamics is the optimization of thermodynamic efficiency of a molecular-scale machine which performs useful work without excessive dissipation, using thermodynamic length \cite{j1,Ruppeiner,k,l,m}. Further extensions of the concept of thermodynamic length for microscopic systems involving the Fisher information matrix can be found in refs \cite{crooks1,crooks2}. Recently, Sivak and Crooks formulated a linear response framework for optimal protocols that minimize dissipation during non-equilibrium perturbations of microscopic systems \cite{crooks3}.

 \smallskip

 A particularly interesting aspect of non-equilibrium thermodynamics ensues while considering systems arbitrarily far from equilibrium (not necessarily in the linear response regime) but working under steady state conditions. Such thermodynamic states are known as non-equilibrium steady states (NESS) \cite{Groot, Bruch} and primarily differ from thermodynamic equilibrium states in respect to the fact that there is a non-zero entropy production since they are not equilibrium states. However, the steady nature of the problem implies that entropy is being accumulated at a constant rate. There has been a notably large body of work in the field of NESS recently \cite{a,b,c,Groot,Gyarmati,Lebon,cohen}. The primary aim of this paper is to develop a geometric formulation of non-equilibrium steady states in quantum thermodynamics in parallel to the developments in equilibrium thermodynamics of systems in the thermodynamic limit.

\smallskip

\textbf{Motivation and plan:} Since several decades, contact geometry has been considered to be a suitable framework for the geometric formulation for dissipative mechanical systems and also for thermodynamics \cite{con1,con2,con21,con3,con4} with varied motivations \cite{mot1,mot2,mot3}. However, most of the developments are confined to the field of equilibrium thermodynamics with thermodynamic transformations corresponding to contact Hamiltonian flows on the Legendre submanifolds \cite{con1,con2,mot2,mot3}. Although there are formalisms such as the general equation for non-equilibrium reversible-irreversible coupling (GENERIC) for dealing with dynamical irreversible situations \cite{generic1,generic2,generic3}, these need to be developed further for applications. The GENERIC, which describes a general class of dynamics obtained by pasting a symmetric dissipative part to an anti-symmetric Poisson part has been studied from the point of view of contact geometry. It should be pointed out that the GENERIC admits a natural geometric description as a second class metriplectic system \cite{metriplectic,metriplectic1}. However, such directions of work are different from the point of view adapted in the present paper where the formalism for a class of quantum steady states such as steady state particle transport through a quantum dot closely following the formalism in \cite{taniguchi} is developed in parallel to that known for equilibrium thermodynamics. It will be shown that the non-equilibrium thermodynamic phase space has an underlying contact structure associated with it and contact Hamiltonian dynamics can be used to describe transformations between different steady states. We also develop a Hamilton--Jacobi theory where the steady state extension of the Massieu--Planck function takes the role of the principal function. Finally, for systems with exponential reduced density matrices it is shown that the thermodynamic metric (equivalent to the Fisher information matrix) on control parameter spaces as presented in \cite{Guarnieri,l} smoothly comes out from our formalism and that other thermodynamic Hessian metrics on control parameter spaces are conformally equivalent to it. This firmly establishes the general nature of the Fisher information matrix and also provides other equivalent ways in which it might be computed.

\smallskip

With this background, the rest of the paper is organized as follows. In the next section, we discuss some basic aspects of steady state quantum thermodynamics of a class of nanoscale systems. We then provide a minimal digression on contact geometry which shall make the paper self contained. Section-(III) is devoted to our results on the geometry of quantum NESS. Finally, we shall end with discussions on our results in section-(IV).

\section{Basics of NESS \& Contact Geometry}\label{section2}
We shall start by briefly describing the notion of steady state quantum thermodynamics of nanoscale systems with a simple example and also set up our notation. This is done in the subsection below.
%One may ask a basic question: what is the difference between a set of reactions in a test tube and a living cell? While the reactions occurring in a test tube is an example of a closed system with chemical isolation, the latter is open to exchange both the chemical energy and materials with its environment. If this exchange phenomenon is sustained for a long time, then an open system usually approaches a steady state that is not an equilibrium state and is known as a non-equilibrium steady state (NESS).
In the subsequent subsection, we describe the basics of contact geometry and Hamiltonian dynamics on contact manifolds.

\subsection{Quantum thermodynamics of nanoscale steady states}
In this subsection, we review some aspects of steady state quantum thermodynamics of a specific class of nanoscale systems (see for example \cite{Guarnieri,taniguchi}) based on which our structure is inspired. We thus consider a nanoscale system such as a quantum dot in contact with the environment. This contact with external baths essentially makes the evolution of the system dissipative or non-unitary. If the system-bath interaction is sufficiently weak then the dynamical description is Markovian and the density matrix satisfies a Lindblad form of master equation. For a nanoscale system, this is typically not the case since the system couples to the reservoirs rather strongly \cite{taniguchi}. The external baths impose upon the system certain boundary conditions such as temperatures, chemical and electric potentials leading to the notion of what are known as the control parameters. It is often convenient to define control parameters \(\{\lambda^i\}\) as some combinations of the external conditions. These control parameters are essentially the variables one is able to manipulate externally, say in an experiment.
\smallskip
\begin{figure}[t]
\begin{center}
\includegraphics[scale=0.7]{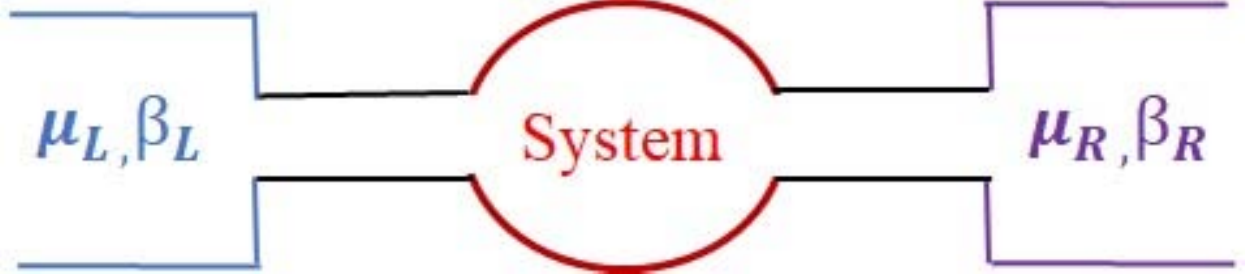}
\caption{Schematic diagram of a two-terminal model set up.}
\label{fig1}
\end{center}
\end{figure}
For example, consider the typical scenario of a two terminal system as shown in figure-(\ref{fig1}) which consists of a central quantum dot in contact with two baths (\(L\) and \(R\)) respectively at (inverse) temperatures \(\beta_L\) and \(\beta_R\). Further, the chemical potentials of the baths are \(\mu_L\) and \(\mu_R\) respectively. A more general setup can be considered, as shown earlier by Taniguchi in \cite{taniguchi}. Had it been that \(\beta_L = \beta_R\) and \(\mu_L = \mu_R\), the system's steady state distribution is the grand canonical distribution. However, in the general case with \(\beta_L \neq \beta_R\) and \(\mu_L \neq \mu_R\) there would be both energy and particle transport taking place across the central quantum dot. A suitable combination of the externally imposed intensive variables can form the set of control parameters. An essential difference between the framework of equilibrium thermodynamics and that of the quantum non-equilibrium steady states being considered here is that in the latter, finding a correct and consistent choice of control parameters is quite non-trivial. An important choice consistent with that adapted in \cite{taniguchi} is to first define the local parameters,
\begin{equation}\label{tempchem}
  \overline{\beta} = \frac{\gamma_L \beta_L + \gamma_R \beta_R}{\gamma_L + \gamma_R}, \hspace{2mm} \overline{\beta \mu} = \frac{\gamma_L \beta_L \mu_L + \gamma_R \beta_R \mu_R}{\gamma_L + \gamma_R}
\end{equation} where \(\gamma_L\) and \(\gamma_R\) are the individual hybridization strengths of the two reservoirs with the system. The set \((\overline{\beta},\overline{\beta \mu})\) therefore defines the average of the external conditions whose thermodynamic conjugates turn out to be the average energy and average particle number of the system in the steady state \cite{taniguchi}. In addition, one defines the two affinities (in a sense, generalized gradients) associated with each reservoir as,
\begin{equation}\label{gradientparameters}
  A_\alpha^E = \overline{\beta} - \beta_\alpha, \hspace{2mm} A_\alpha^N = -(\overline{\beta \mu} - \beta_\alpha \mu_\alpha)
\end{equation} where \(\alpha = L, R\). Clearly, if \(\beta_L \neq \beta_R\) and \(\beta_ L\mu_L \neq \beta_R \mu_R\), the affinities are non-vanishing and there is a non-zero energy and particle transport through the system. However, if the control parameters are held fixed, the system approaches a non-equilibrium steady state with a fixed average energy and average number of particles. Such steady states are characterized by a steady (time independent) yet positive definite rate of entropy production, \(\Sigma > 0\). The system approaches a local equilibrium state (LES) if all the external gradients vanish (\(A_\alpha^E=A_\alpha^N=0\)) meaning that an arbitrarily small volume of the system is effectively homogeneous i.e. without an effective spatial structure. Generically, the control parameters are associated with conjugate response functions \(\{F_i\}\) which variables which may depend on the system size and include the currents and fluxes driven by the external conditions. Even though in the non-equilibrium regime there are complicated fluctuations associated with the response functions which may be of thermal or quantum origin or both, in a stationary or steady state, the averages values \(\langle F_i \rangle = X_i\) can be taken to describe the system. Note that the currents may in general be nonlinear. For the central system (a quantum dot) shown in figure-(\ref{fig1}), within the wide-band approximation the net currents through the system are given by the Landauer--B\"{u}ttiker formula \cite{LB1,LB2,LB3},
\begin{eqnarray}
% \nonumber % Remove numbering (before each equation)
  J_N &=& \frac{2}{\hbar}\frac{\gamma_L\gamma_R}{\gamma}\int d\epsilon \rho(\epsilon)[f_L(\epsilon)-f_R(\epsilon)] ,\\
  J_E &=& \frac{2}{\hbar}\frac{\gamma_L\gamma_R}{\gamma}\int d\epsilon \rho(\epsilon)\epsilon[f_L(\epsilon)-f_R(\epsilon)],
\end{eqnarray}
where $\rho(\epsilon)=\frac{1}{\pi}\frac{\gamma}{(\epsilon-\epsilon_s)^2+\gamma^2}$, $\epsilon_s=$ system energy, $\gamma=\gamma_L+\gamma_R$ while $f_{L,R}=[e^{\beta_{L,R}(\epsilon-\mu_{L,R})}\mp 1]^{-1}$ are the distribution functions of the reservoirs. A key ingredient in our analysis will be the existence of a steady state Massieu--Planck function (or Massieu potential) \(\psi_{ness}\) whose significance can be expressed as (see for example the recent work \cite{taniguchi}),
\begin{equation}
  d\psi_{ness} = E  d\overline{\beta} + N  d\overline{\beta \mu} + \sum_\alpha \bigg( J_{\alpha E} dA_{\alpha}^E +  J_{\alpha N} dA_{\alpha}^N \bigg)
\end{equation} where we have set \(\hbar = 1\) and have absorbed all constant factors into the expressions for \(J_{\alpha E}\) and \(J_{\alpha N}\). Note that \(J_{\alpha E}\) and \(J_{\alpha N}\) are the energy and particle currents associated with the individual reservoirs. The Massieu potential is time independent and stays constant in any steady state. Still, it is capable of characterizing the steady state thermodynamic characteristics and the entropy production rate \cite{Guarnieri,taniguchi}. In case of \(n\) arbitrary control parameters one has,
\begin{equation}\label{firstlaw}
  d\psi_{ness} - X_i  d\lambda^i=0.
\end{equation}
Clearly, the Massieu potential is a function of the control parameters, i.e. \(\psi_{ness} = \psi_{ness}(\lambda^i)\). In these scenarios, it is often possible to define a reduced density matrix \(\rho\) describing the steady state characteristics of the system in the McLennan--Zubarev form \cite{McLennan,Zubarev,Hershfield,Ness}. Then the Massieu potential may directly be found out as, \(\psi_{ness} = \ln Z_{ness}\) where \(Z_{ness}\) is the associated NESS partition function \cite{Guarnieri,taniguchi}. It should be specially emphasized that since such a reduced density matrix is time independent because it describes only the steady state characteristics, the Massieu potential naturally turns out to be time independent in contrast with the entropy which is being constantly accumulated with a steady rate. Remarkably, the exact functional form of \(\psi_{ness}\) shall not be relevant for our analysis in this paper as long as it is at least twice differentiable in \(\{\lambda^i\}\) and satisfies eqn (\ref{firstlaw}) for steady states. This ensures that our results are quite robust within the scope of systems studied in \cite{taniguchi}. However, since eqn (\ref{firstlaw}) is the starting point for our analysis, it is specifically tailored for systems where an analogous statement (such as eqn (\ref{firstlaw})) holds good. This shall therefore not include all non-equilibrium quantum systems working under steady state conditions but a specific class of them whose properties were investigated in some detail in \cite{taniguchi}. Having said that, eqn (\ref{firstlaw}) allows one to write the response variables as,
\begin{equation}
   X_i  = \frac{\partial \psi_{ness}}{\partial \lambda^i}.
\end{equation}
In this sense, eqn (\ref{firstlaw}) is the first law for the steady state quantum thermodynamics of nanoscale systems and is quite general because it does not require thermodynamic equilibrium. We remark, after having made this identification that the second law of NESS thermodynamics is simply the statement \(\Sigma > 0\) \cite{cohen}.

\smallskip

 Finally, let us note that the control parameters \(\{\lambda^i\}\) being smoothly varying can without any loss of generality be thought of as local coordinates in some open subset of \(\Re^n\) with \(n\) being the number of control parameters present to manipulate the system. It shall be assumed that these parameters are all independent of each other and that there are no constraints between them. For the system considered in figure-(\ref{fig1}), this means that one can change the temperature of one of the baths (say \(\beta_L\)) without having to worry about the \(\beta_R,\mu_L\) or \(\mu_R\) which remain unaffected. In that case, the control parameters can be considered to be local coordinates on some \(n\)-dimensional smooth manifold \(Q\) which we shall call the control parameter space for the system.

%By definition, the entropy \(S\) being an extensive thermodynamic variable is only a function of the extensive response functions. We shall particularly focus on systems where the entropy \(S=S( X_i )\) has the structure of a differentiable function. The conjugate control parameters are then given as,
%\begin{equation}
 % \lambda^i = \frac{\partial S}{\partial X_i },
%\end{equation} which can be re-expressed as,
%\begin{equation}\label{dS}
 % dS - \lambda^i d  X_i  = 0.
%\end{equation}
%Since it is the set of variables \(\{\lambda^i\}\) which one controls externally, it is desirable to have a function of these control parameters replacing \(S\) from the previous two equations. Such a function known as the Massieu potential or the free entropy and is defined as the Legendre transformation of the entropy,
%\begin{equation}\label{M}
 % \psi_{ness} = S - \lambda^i X_i .
%\end{equation}
%It should be pointed out that a system admits a description in terms of the Massieu potential only if the entropy is a convex function so that the Legendre transformation [eqn (\ref{M})] is meaningful. In terms of Massieu potential one can write,
%\begin{equation}\label{firstlaw}
 % d\psi_{ness} = - X_i  d\lambda^i ,
%\end{equation}
%where \(\psi_{ness}=\psi_{ness}(\lambda^i)\) is a function of the control parameters only. This equivalently means that \(\psi_{ness}:Q \rightarrow \Re\), i.e. the Massieu potential is defined on the control manifold of the system.

\subsection{Elements of contact geometry}
Contact geometry \cite{Geiges, Arnold} is the odd dimensional counterpart of symplectic geometry \cite{S} which forms the natural geometric setting for classical Hamiltonian mechanics. Hamiltonian dynamics can be formulated on contact manifolds and this dynamics naturally incorporates a dissipative character \cite{CM1, CM2}. Contact geometry has further been applied quite extensively to reversible and irreversible thermodynamics (see for example \cite{con3, RT3, RT2, generic1, generic2, generic3} for some notable works) as well as to statistical mechanics \cite{SM}. We shall for the sake of completeness briefly recap some elements of contact geometry in this section. The central concept is that of a contact manifold which is the pair \((\mathcal{M},\eta)\) where \(\mathcal{M}\) is a \((2n+1)\)-dimensional smooth manifold and \(\eta\) is a one form that satisfies the condition of complete non-integrability,
\begin{equation}\label{nonint}
  \eta \wedge (d\eta)^n \neq 0 .
\end{equation}
More explicitly, the kernel of the one form \(\eta\) defines a distribution of hyperplanes which is completely non-integrable in the Frobenius sense. Indeed, non-integrability in the sense of Frobenius would mean \(\eta \wedge d\eta \neq 0\) and eqn (\ref{nonint}) is therefore a much stronger statement. This condition of complete non-integrability implies that the tangent bundle of \(\mathcal{M}\) can be decomposed into the following Whitney sum,
\begin{equation}
  T\mathcal{M} = ker(\eta) \oplus ker(d\eta)
\end{equation} where \(ker(\eta)\) and \(ker(d\eta)\) are both regular distributions. Moreover, there exists a unique global vector field \(\xi\), known as the Reeb vector field defined through,
\begin{equation}\label{reeb}
  \eta(\xi) = 1, \hspace{3mm} d\eta(\xi,.)=0
\end{equation} meaning that the flow of the Reeb vector field \(\xi\) preserves \(\eta\) and consequently the hyperplane distribution defined by \(\ker(\eta)\). It is always possible to find local (Darboux) coordinates in the neighbourhood of any point on \(\mathcal{M}\) and in such a Darboux chart \((s,q^i,p_i)\) with \(i=1,2,....,n\), the one form \(\eta\) and the Reeb vector field \(\xi\) are expressed as,
\begin{equation}
  \eta = ds - p_idq^i, \hspace{3mm} \xi = \frac{\partial}{\partial s} .
\end{equation}
A simple computation can verify eqns (\ref{reeb}). It is interesting to note that,
\begin{equation}
  d\eta = dq^i \wedge dp_i
\end{equation} which is the local expression of the standard symplectic form \(\omega\) on any \(2n\)-dimensional symplectic manifold (see for example \cite{S}). It is then not hard to convince oneself that a \((2n+1)\)-dimensional contact manifold \((\mathcal{M},\eta)\) has a \(2n\)-dimensional symplectic submanifold where \(\omega=d\eta\) is the symplectic two form. Further, since \(d\eta(\xi,.)=0\) identically, the Reeb vector field is inconsequential on this symplectic submanifold.

\subsubsection{Contact Hamiltonian dynamics}
Analogous to the Hamiltonian dynamics formulated on symplectic manifolds, one can also formulate Hamiltonian dynamics on a contact manifold \((\mathcal{M},\eta)\) as follows. For an arbitrary smooth function \(h:\mathcal{M} \rightarrow \Re\), one associates a vector field \(X_h\), also called the contact vector field obtained from \(h\) through the following conditions,
\begin{equation}
  \eta(X_h) = -h, \hspace{3mm} i_{X_h}d\eta = dh - \xi(h)\eta .
\end{equation}
In a local Darboux chart, the contact vector field \(X_h\) takes the form,
\begin{equation}\label{contactfield}
  X_h = \bigg(p_i\frac{\partial h}{\partial p_i}-h\bigg)\frac{\partial}{\partial s} - \bigg(p_i \frac{\partial h}{\partial s}+\frac{\partial h}{\partial q^i}\bigg)\frac{\partial}{\partial p_i} + \bigg(\frac{\partial h}{\partial p_i}\bigg)\frac{\partial}{\partial q^i} \,
\end{equation}which means that the evolution equations have the form,
\begin{equation}\label{eqnsofmotion}
  \dot{s} =  p_i\frac{\partial h}{\partial p_i} - h; \hspace{3mm} \dot{q}^i = \frac{\partial h}{\partial p_i}; \hspace{3mm} \dot{p}_i = -p_i \frac{\partial h}{\partial s} - \frac{\partial h}{\partial q^i} \, .
\end{equation}
The resemblance with the classical Hamilton's equations is clear. This dynamics can be shown to correspond to thermodynamic transformations in equilibrium thermodynamics where the local coordinates \((s,q^i,p_i)\) are identified as coordinates in the thermodynamic phase space. Furthermore, a contact manifold is an example of a Jacobi manifold, i.e. it is endowed with a bilinear local Lie bracket structure \cite{Jacobi, Jacobi2}. The commutator of two contact vector fields defines an anti-symmetric bilinear bracket between the functions generating them. Such a bracket has been called a Lagrange bracket \cite{Rajeev, Ghosh} and is given by,
\begin{equation}
\{q^i,p_j\} = \delta^i_j .
\end{equation}
The Leibniz rule is however not obeyed by the Lagrange bracket and therefore the Lagrange bracket of a function with a constant function does not vanish in general. The coordinates \(q^i\) and \(p_i\) satisfying a non-trivial Lagrange bracket relationship are identified as conjugate variables. The equations of motion [eqns (\ref{eqnsofmotion})] are also clearly dissipative in a mechanical context since the equation of motion for \(p_i\) includes a friction term with the damping factor being given by \(\xi(h)\). Unlike the symplectic case, the function \(h = h(s,q^i,p_i)\) is not conserved along the dynamics, i.e. \(X_h(h) \neq 0\) whenever \(h \neq 0\) (or \(h\) is independent of \(s\)). In fact, the additional coordinate \(s\) models an interaction of the system with some environment leading to nice thermostat problems \cite{SM}. A straightforward calculation can reveal that one can write down the dynamics of an arbitrary smooth function \(f:\mathcal{M} \rightarrow \Re\) directly using the Lagrange bracket as,
\begin{equation}
  \dot{f} = - h \xi(f) + \{f,h\}.
\end{equation}
Note that if \(s\) does not appear in \(h\), the equations of motion are simply the usual Hamilton's equations with \(s\) being the action functional. The dynamics is then purely conservative. Due to the resemblance of the dynamics generated by \(h\) with the classical Hamiltonian dynamics, the function \(h\) is called a contact Hamiltonian while the triplet \((\mathcal{M},\eta,h)\) is called a contact Hamiltonian system.

\subsubsection{Legendre submanifolds}
Of interest particularly in thermodynamics are a very specific class of submanifolds of a contact manifold. They are the submanifolds of the maximum possible dimension such that they do not include a conjugate pair of local coordinates. In more technical terms, if \(L \subset \mathcal{M}\) be such a submanifold, then all the tangent vectors of \(L\) are annihilated by \(\eta\) or in other words, restriction of the contact form \(\eta\) should be zero on the submanifold. These submanifolds are known as Legendre submanifolds and are equivalent to configuration spaces from classical mechanics. The contact form \(\eta\) when restricted to an arbitrary Legendre submanifold \(L\) vanishes, i.e. in local coordinates,
\begin{equation}\label{contactvanish}
  [ds - p_idq^i]_L = 0
\end{equation} whence it is clear that \(L\) does not include any conjugate pair of variables also meaning that the Lagrange bracket between any two independent coordinates on \(L\) is zero identically. It follows that all Legendre submanifolds are \(n\)-dimensional and their local structure is given by \cite{S},
\begin{equation}\label{legendrelocal}
   p_i = \frac{\partial F}{\partial q^i}, \hspace{3mm} q^j = -\frac{\partial F}{\partial p_j},  \hspace{3mm} s = F - p_j \frac{\partial F}{\partial p_j} \,
\end{equation} where \(i \in I, j \in J\) and \(I \cup J\) is a disjoint partition of the set of \(n\) indices. The function \(F=F(q^i,p_j)\) is known as the generator of Legendre submanifold. In the next section, we shall see that the Massieu potential typically plays the role of such a generator function.

\subsubsection{Metric structures on a contact manifold}
It was shown long back that contact manifolds in general can be associated with metric structures that are compatible with the contact structure (see \cite{metric1,metric2,metric3} for the details). Such a compatible metric \(G\) is a bilinear, symmetric and non-degenerate function, \(G:T\mathcal{M} \times T\mathcal{M} \rightarrow \Re\) where \(T\mathcal{M}\) is the tangent bundle of \(\mathcal{M}\). Taking the metric to be of the form \( G = \eta^2 + dp_idq^i\) (which is seen to be bilinear, symmetric as well as non-degenerate) it follows that on a Legendre submanifold \(L\) with a local form given by eqn (\ref{legendrelocal}), it takes the form,
\begin{equation}\label{metriclocal}
   G_L= dp_idq^i|_{L} = \frac{\partial^2 F}{\partial q^i \partial q^{i'}}dq^idq^{i'} - \frac{\partial^2 F}{\partial p_j \partial p_{j'}}dp_jdp_{j'}
\end{equation} where \(i,i' \in I\) and \(j,j' \in J\) with \(I \cup J = \{1,2,....,n\}\) being a disjoint partition of the index set. This metric allows one to define the notion of a length between two points, i.e. two different configurations on the Legendre submanifold \(L\). As will be shown next, the NESS control parameter space shall assume the structure of a Legendre submanifold and such a metric shall then define the distance between two control parameter configurations.

\section{Geometry of quantum NESS: Some results}
We will now show that the non-equilibrium steady states correspond to points on Legendre submanifolds of the non-equilibrium thermodynamic phase space which has the structure of a contact manifold. In this context we propose several results which will further the understanding of the geometric structure of NESS. Recalling the steady state relation \(d\psi_{ness} - X_i d\lambda^i = 0\) immediately leads to the identification that \((\psi_{ness},\lambda^i, X_i )\) are local coordinates on a contact manifold \((\mathcal{M},\eta)\) with \(\eta = d\psi_{ness} - X_i d\lambda^i\). The steady state condition [eqn (\ref{firstlaw})] demands \(\eta = 0\) and therefore a non-equilibrium steady state is simply a point on an appropriate Legendre submanifold with the local structure being given as,
\begin{equation}\label{legendrelocalNESS}
 s = \psi_{ness} (\lambda^i), \hspace{3mm} X_i = \frac{\partial \psi_{ness}(\lambda^i)}{\partial \lambda^i} .
\end{equation}
Since any contact manifold of \((2n+1)\)-dimensions can locally be thought of as \(T^*\Re^n \times \Re\) where \(T^*\Re^n\) is the cotangent bundle of \(\Re^n\), it follows that \(\{\lambda^i\}\) are the coordinates in \(\Re^n\) while \(\{ X_i\}\) are the cotangent fiber coordinates. The Legendre submanifold is locally an open set in \(\Re^n\) with the local coordinates being the external control variables. Based on the assumptions made in section-(1) on the systems considered in this work, the following proposal can be made,
\begin{prop}\label{Lemma1}
 A set of \(n\) independent parameters \(\{\lambda^i\}\) being local coordinates in an open subset of \(\Re^n\) together with a smooth function \(\psi_{ness}:\Re^n \rightarrow \Re\) (the Massieu potential) constitutes a non-equilibrium quantum system in steady state with corresponding steady state response functions or thermodynamic forces being given by,
\begin{equation}\label{12345}
 X_i = \frac{\partial \psi_{ness}(\lambda^i)}{\partial \lambda^i}.
 \end{equation}
\end{prop}

 As also stated earlier, the parameter space \(Q\) has the structure of a smooth manifold and is therefore locally equivalent to \(\Re^n\). In fact, what is not so trivial is that any such system can be lifted (at least locally) to a space with an ambient contact structure as has been stated as follows,

\begin{theorem}\label{theorem1}
  A non-equilibrium quantum system \((\lambda^i,\psi_{ness}(\lambda^i))\) with an \(n\)-dimensional control parameter space \(Q\) can be lifted to a non-equilibrium thermodynamic phase space which is a contact manifold \((\mathcal{M},\eta)\) of \((2n+1)\)-dimensions via an inclusion map \(\Phi:Q \rightarrow \mathcal{M}\) such that \(\Phi^*\eta = 0\) where \(\Phi^*\) is the pullback induced by \(\Phi\).
\end{theorem}

This result can be understood as follows. Given that one has a quantum system in steady state \((\lambda^i,\psi_{ness}(\lambda^i))\) with a parameter space or control parameter space \(Q\), the space \(T^*Q \times \Re\) is always locally isomorphic to \(T^*\Re^n \times \Re\) on which one has the standard contact form \(\eta\) constructed locally as \(\eta = ds + d\theta\) where \(d\theta = -  X_i  d\lambda^i\) is the tautological one form on \(T^*Q \approxeq T^*\Re^n\) and \(s \in \Re\). On \(Q\) one has \(s = \psi_{ness}(\lambda^i)\) so that \( X_i \), being the fiber coordinates on \(T^*Q\) are given by eqn (\ref{12345}) (also see \cite{quevedo} and references therein).

\smallskip

Note that different points on \(Q\) correspond to different values of the control parameters and hence different non-equilibrium steady states. In \cite{Guarnieri}, the parameter space \(Q\) has been called the manifold of steady states. By theorem-(\ref{theorem1}) the control parameter space can be thought of as being embedded in an ambient thermodynamic phase space which assumes the structure of a contact manifold where the following Lagrange bracket relationship holds,
\begin{equation}
  \{\lambda^i,  X_j\} = \delta^i_j .
\end{equation}
Such relations may prove useful in introducing thermodynamic uncertainty relations (see for example \cite{Rajeev2}).

\subsection{Transformations between steady states}
In this subsection, we shall provide a Hamiltonian description of transformations between steady states on \(Q\). By changing the external control parameters, it is possible to drive the system into a different steady state. For instance, if the temperatures of two reservoirs connected at two ends of a system are changed, it is possible to change the heat current flowing through the system. But arbitrary changes to the control variables may drive the system into an unsteady state. For this reason, it will be assumed throughout that the transformations are performed quasi-statically so that between any two initial and final points on \(Q\), the system passes through an infinitely many steady states at an infinitesimally small speed. Clearly however, for an initial steady state and a final one, there are infinitely many paths connecting the two. In order to pick up the optimal transformation path, we assume the existence of a friction tensor which in the Kirkwood formalism can be derived as the time integral of the force covariance matrix \cite{friction1,friction2}.
\begin{prop}
  For a non-equilibrium quantum system \((\lambda^i,\psi_{ness}(\lambda^i))\) in steady state with a control parameter space \(Q\), it is possible to associate an \(n \times n\) positive definite matrix \(M_{ij}\), known as the friction tensor whose components are functions of the control parameters, i.e. \(M_{ij}=M_{ij}(\lambda^i)\).
\end{prop}

 %be using contact Hamiltonian dynamics to describe the optimal paths on a control manifold. Within the context of several non-equilibrium systems working in steady states such as molecular machines, one desires to find an optimal path on the control parameter space such that along this particular path which joins two end points, the system undergoes minimum dissipation as compared to all other paths connecting the same two points! In ref \cite{crooks3} working in the linear response regime, the authors have associated the notion of a friction tensor \(M_{ij}\) which is a positive semi-definite symmetric matrix defined over the control manifold such that the optimal paths are geodesics of this friction tensor.

Having defined the friction tensor as a positive definite matrix, we consider following contact Hamiltonian function on the thermodynamic phase space,
\begin{equation}\label{contactHamiltonianoptimal}
    h = M^{ij}\bigg( X_i X_j - \frac{\partial \psi_{ness}}{\partial \lambda^i}\frac{\partial \psi_{ness}}{\partial \lambda^j}\bigg)
  \end{equation} which is one choice of contact Hamiltonian describing quasi-static transformations between two steady states of a non-equilibrium quantum system where \(M^{ij}\) is the inverse friction tensor \footnote{A careful reader would have noticed that the friction tensor \(M_{ij}=M_{ij}(\lambda^i)\) is defined only over the control parameter space whereas the symmetric matrix (actually its inverse) appearing in the expression of the contact Hamiltonian [eqn (\ref{contactHamiltonianoptimal})] should be defined over the entire thermodynamic phase space. We have been slightly abusive with our notation in labelling both of them with the same symbol. The friction tensor on the control parameter space \(Q\) is simply the full symmetric matrix on \(\mathcal{M}\) pulled back to \(Q\), i.e. \(\Phi^*M\) where \(\Phi: Q \rightarrow \mathcal{M}\) is the relevant inclusion map.}.

\smallskip

One can obtain the contact vector field \(X_h\) corresponding to this choice of the Hamiltonian [eqn (\ref{contactfield})] (of course with the usual identification \((\lambda^i,X_i) \rightarrow (q^i,p_i)\)). Recall that one has \(X_h(h) \neq 0\) unless \(h = 0\) identically and therefore \(X_h\) is not in general tangent to the level surfaces of constant \(h\). However, for the present case we note that by eqn (\ref{12345}), one has \(h|_Q = 0\) and therefore the dynamical flow is tangent to \(Q\). We are interested in the dynamics of the control parameters and hence projecting \(X_h\) to \(Q\) one finds,
\begin{equation}\label{26}
  X_h|_Q = M^{ij} \frac{\partial \psi_{ness}}{\partial \lambda^j} \frac{\partial}{\partial \lambda^i} \,
\end{equation} where we have used eqn (\ref{12345}). This means that the dynamics of any smooth function \(f\) on the control parameter space, i.e. \(X_h|_Q(f)\) is given by,
\begin{equation}\label{12345678}
   \frac{d f}{d \tau} = M^{ij}  \frac{\partial f}{\partial \lambda^i} \frac{\partial \psi_{ness}}{\partial \lambda^j}
 \end{equation} where \(\tau \in \Re\) is an affine parameter which parameterizes the trajectory. At the first look, this resembles the dynamics of a function from the standard Hamiltonian scenario but now with a symmetric matrix replacing the anti-symmetric Poisson matrix and \(\psi_{ness}\) playing the role of the function which generates the dynamics. Note that we are now entirely on the control parameter space \(Q\) and not the full thermodynamic phase space whose details have become unimportant at this stage. Introducing a symmetric bracket or in other words an inner product as,
 \begin{equation}\label{innerproduct}
   (f,g) = M^{ij} \frac{\partial f}{\partial \lambda^i} \frac{\partial g}{\partial \lambda^j}
 \end{equation} for any two smooth functions \(f,g:Q \rightarrow \Re\), one can re-express eqn (\ref{12345678}) as follows,
 \begin{equation}
   \frac{d f}{d \tau} = (f,\psi_{ness}).
 \end{equation}
This is in some sense a Poisson analogue defined for the present non-equilibrium regime. However, unlike the Poisson case where the standard Hamiltonian generating the dynamics stays constant along the flow, in this case \(\psi_{ness}\) is not preserved along the dynamics. In fact one has,
\begin{equation}\label{mnb}
    \frac{d \psi_{ness}}{d \tau} = (\psi_{ness},\psi_{ness})
\end{equation} which is obviously non-zero unless \(\psi_{ness} = 0\). The following result can now be stated,
\begin{theorem}\label{theorem3}
  The dynamics described by the contact Hamiltonian suggested in eqn (\ref{contactHamiltonianoptimal}) is such that when projected on the control parameter space \(Q\), the Massieu potential (free entropy) always increases along the transformation.
\end{theorem}
This follows simply from the fact that for any non-zero smooth function \(f:Q \rightarrow \Re\), one has \((f,f) > 0\) for the bracket defined in eqn (\ref{innerproduct}). Theorem-(\ref{theorem3}) is consistent with the fact that a transformation between two non-equilibrium steady states implies that the Massieu potential \(\psi_{ness}\) which can be interpreted as the free entropy increases along the transformation. In other words, a quasi-static transformation between two steady states implies \(\Delta \psi_{ness} > 0\).

\smallskip

We now proceed forward to check whether the formalism we have been describing indicates towards the fact that the geodesics of the friction tensor are the optimal paths. For quasi-static transformations between steady states, such paths are those for which the steady state Massieu potential changes the least as compared to any other path. Now consider two different paths \(\gamma\) and \(\gamma'\) between the same end points in \(Q\) and let \(||\gamma'|| > ||\gamma||\) where \(||.||\) denotes the length of the particular path. Also let \(\gamma\) be the geodesic of \(M^{ij}\) between the end points. Surely then, \(\gamma'\) is not a geodesic of \(M^{ij}\). It has a length greater than that of \(\gamma\) and therefore can generically be thought of as the geodesic due to some symmetric matrix \(M^{ij} + \delta M^{ij}\) where \(\delta M^{ij}\) is another positive definite symmetric matrix such that if the system is to traverse along \(\gamma'\) the dynamics is to be defined as,
\begin{equation}
  \frac{d f}{d \tau}\bigg|_{M + \delta M} = (M^{ij} + \delta M^{ij}) \frac{\partial f}{\partial \lambda^i} \frac{\partial \psi_{ness}}{\partial \lambda^j}
\end{equation} where the subscript \(M + \delta M\) denotes the fact that the path taken is not the geodesic of the friction tensor \(M\) but that of the symmetric matrix \(M + \delta M\) (dropping indices for the moment just for the sake of brevity). It is not hard to see that in this case one has,
\begin{equation}
  \frac{d \psi_{ness}}{d \tau}\bigg|_{M + \delta M} - \frac{d \psi_{ness}}{d \tau}\bigg|_M > 0.
\end{equation}
Since, \(\delta M\) is arbitrary we may conclude from here that the (free) entropy production along the geodesic of the friction tensor \(M^{ij}\) is the minimum, i.e. one can say,
\begin{theorem}\label{thhh}
  On a control parameter space \(Q\) equipped with a symmetric friction tensor \(M^{ij}\), the optimal paths of quasi-static transformations between steady states are the geodesics of \(M^{ij}\).
\end{theorem}

We note that our contact Hamiltonian [eqn (\ref{contactHamiltonianoptimal})] partly resembles that proposed in \cite{generic2} within the context of the general equation for non-equilibrium reversible-irreversible coupling (GENERIC). However, in the present case, the dynamics on the control parameter space lacks a Poisson part (unlike the case with GENERIC) making it purely of symmetric nature.

\subsection{Steady state Massieu potential as a principal function}
In this section, we shall strengthen the analogy between the physics of steady states in quantum systems and mechanics. To this end, let us define an extended non-equilibrium thermodynamic phase space \(\mathcal{M}' = \mathcal{M} \times \Re\) with \(\mathcal{M}\) being the non-equilibrium thermodynamic phase space. If \(\eta\) is the contact one form on \(\mathcal{M}\), one can define a Poincare form on \(\mathcal{M}'\) as, \(\eta' = h d\tau + \eta\) where \(h\) is any smooth function on \(\mathcal{M}'\) and \(\tau \in \Re\). We wish to construct a Hamilton--Jacobi theory for the present scenario which for generic contact Hamiltonian systems was discussed earlier in \cite{CM2}. In doing so, for the principal function \(W(\lambda^i,\tau)\) defined on \(Q \times \Re\) we seek for solutions to the partial differential equation,
\begin{equation}
  \mathcal{F}\bigg( \lambda^i, \tau, W, \frac{\partial W}{\partial \lambda^i}, \frac{\partial W}{\partial \tau} \bigg) = 0
\end{equation} whose characteristic curves are equivalent to quasi-static transformations between steady states. Let us define, \(\mathcal{F} = E - h\) where \(h\) is given by eqn (\ref{contactHamiltonianoptimal}) and \(E\) is a constant. The condition \(\mathcal{F} = 0\) then implies that \(h\) must be equal to the constant \(E\) on a subspace \(Q_E\) parameterized by \(E\). It can be shown \cite{CM2} that this is equivalent to having the Poincare one form \([\eta + h d\tau]_{Q_E = 0}\) on the surface \(Q_E\) parameterized by constant \(E\), i.e.
 \begin{equation}
   [dW - X_i d\lambda^i + h d\tau]_{Q_E} =0
 \end{equation} giving on \(Q_E\) the following equalities,
 \begin{equation}\label{mnbvcxz}
  X^i = \frac{\partial W(\lambda^i,\tau)}{\partial \lambda^i}, \hspace{3mm}  h + \frac{\partial W(\lambda^i,\tau)}{\partial \tau} = 0.
 \end{equation}
This essentially means that under the projection \(\pi: Q \times \Re \rightarrow Q\), the principal function \(W(\lambda^i,\tau)\) goes to the Massieu potential \(\psi_{ness}(\lambda^i)\). In other words, the Massieu potential is equivalent to the Hamilton's principal function under each slice of \(\tau\). If this identification is made, the first among eqns (\ref{mnbvcxz}) is the same as eqn (\ref{12345}). The second one together with the choice of contact Hamiltonian in eqn (\ref{contactHamiltonianoptimal}) is equivalent to the fact that the Massieu potential is explicitly independent of \(\tau\). Thus, one obtains the re-interpretation that the steady state Massieu potential is the analogue of the principal function for steady state quantum thermodynamics. This analogy is fairly robust.

\subsection{Thermodynamic length}
Since the non-equilibrium steady states correspond to points on Legendre submanifolds with the local structure given by eqn (\ref{legendrelocalNESS}), there is a natural Riemannian metric on the control parameter space. On the control parameter space \(Q\), the metric tensor defined in eqn (\ref{metriclocal}) takes the following form,
\begin{equation}\label{metric}
  g_{ij} = \frac{\partial^2 \psi_{ness}}{\partial \lambda^i \partial \lambda^j}, \hspace{5mm} i,j \in \{1,2,....,n\}.
\end{equation} This defines the notion of a thermodynamic length with line element \(dl^2 = g_{ij} d\lambda^i d\lambda^j\) between different points on \(Q\). Note that the metric tensor is defined as the Hessian of the Massieu potential with respect to the control parameters only and we shall call this the Schl\"{o}gl metric \cite{Smetric} (also see \cite{l}).

\smallskip

 Noting that for particular values of the control variables the system reduces to a local equilibrium state, it follows that local equilibrium states are also points on the same control parameter space \(Q\). For example, in the case of the system shown in figure-(\ref{fig1}), local equilibration occurs only when both the baths have the same temperatures and chemical potentials, say \(\beta\) and \(\mu\) respectively. Consequently, local equilibrium states correspond to points on the same control parameter space \(Q\) on which the non-equilibrium steady states lie on. Therefore, our formalism properly generalizes that developed for equilibrium thermodynamics (for example see \cite{con1,con2,con21,con3,con4}). For example, for the system shown in figure-(\ref{fig1}), the control parameter space \(Q\) can be written as, \(Q = Q_{avg} \times Q_{aff}\) where \((\overline{\beta},\overline{\beta \mu})\) are coordinates on the space \(Q_{avg}\) whereas, the affinities \((A_\alpha^E,A_\alpha^N)\) are coordinates on \(Q_{aff}\). This space \(Q_{aff}\) is trivial, i.e. \((A_\alpha^E,A_\alpha^N) = (0,0)\) when the system is at a local equilibrium state. In such a case, just the space \(Q_{avg}\) appropriately describes the system in the grand canonical ensemble with control parameters \(\beta\) and \(\mu\) since the gradients have identically vanished. Thus, \(Q_{avg}\) is the equilibrium thermodynamics configuration space as has been discussed earlier in the literature for various classes of systems. These arguments go through for systems with an arbitrary number of control parameters where it is still possible to decompose the total control parameter space into a configuration space \(Q_{avg}\) and the space of affinities \(Q_{aff}\). The metric [eqn (\ref{metric})] on \(Q\) therefore not only defines the thermodynamic length between different non-equilibrium
 steady states but also includes local equilibrium states.

  \smallskip

  For an affine parameter \(\tau\), the length of a curve between parameter values \(a\) and \(b\) is simply the action,
\begin{equation}
  l(a,b) = \int_{a}^{b} d\tau \sqrt{g_{ij}\frac{d\lambda^i}{d\tau}\frac{d\lambda^j}{d\tau}} .
\end{equation} This formalism may therefore be applied to study geodesics on a control parameter space which represent the shortest distance between two steady states. The notion of congruence of non-intersecting geodesics can be explored further by considering the variations of an appropriate deformation vector that joins two points on geodesics that are close. We emphasize that although there is no general procedure to obtain the probability distribution in the out of equilibrium setting, for several non-equilibrium systems including molecular machines, it is possible to directly compute the density matrix \(\rho\) in the McLennan--Zubarev form which has the structure of a generalized exponential distribution. In that case, the Massieu potential can be obtained as \( \psi_{ness} = \ln Z\) where \(Z\) is the NESS partition function. It should be pointed out here that when the NESS density matrix \(\rho\) has the form of an exponential, a case which is of chief interest here \cite{Guarnieri,taniguchi}, it is possible to associate the metric \(g_{ij}\) defined in eqn (\ref{metric}) with the covariance matrix. This can be seen by writing,
\begin{equation}
  g_{ij} = \frac{\partial X_i}{ \partial \lambda^j} = \frac{\partial \langle F_i \rangle}{ \partial \lambda^j}
\end{equation}
where as also stated earlier, \(X_i\) defines the steady state mean value of the corresponding response function \(F_i\). Thus more explicitly, \(\langle F_i \rangle = {\rm Tr} (\rho F_i)\) (note that the Massieu potential normalizes \(\rho\) meaning that \({\rm Tr} (\rho) = 1\)) and similarly, \(\langle F_i^2 \rangle = {\rm Tr} (\rho F_i^2)\) with \(\rho \sim e^{-\psi_{ness} + \lambda^i F_i}\). It is then not hard to check by explicit calculation that (see also \cite{con21,con4,l}),
\begin{equation}
  g_{ij} = \frac{\partial}{ \partial \lambda^j}{\rm Tr} (\rho F_i)  = \langle F_i F_j \rangle - \langle F_i \rangle \langle F_j \rangle
\end{equation} which implies that the metric tensor coincides with the co-variance matrix of the response functions. Thus, although the metric coefficients would in general include non-equilibrium transport coefficients which are not necessarily positive in fully nonlinear-response regime, the overall metric tensor is essentially positive definite owing to the fact that it represents an appropriate co-variance matrix. Furthermore, this metric \(g_{ij}\) can be shown \cite{Guarnieri,l} to be equivalent to the Fisher information matrix (up to an overall sign) defined directly from the NESS density matrix \(\rho\) as,
\begin{equation}
  g_{ij} = \sum_{k} \rho_k(\lambda^i)\bigg(\frac{\partial \ln \rho_k(\lambda^i)}{\partial \lambda^i}\frac{\partial \ln \rho_k(\lambda^j)}{\partial \lambda^j}\bigg)
\end{equation} where the set \(\{\rho_k\}\) denotes the set of projections of the density matrix on the energy eigenbasis of the full Hamiltonian representing the central quantum system, the baths and relevant interaction terms. However, we should also bear in mind that in a case where the density matrix fails to have an exponential structure, it is not immediately apparent if the metric \(g_{ij}\) coincides with the covariance matrix and is positive definite.

\smallskip

Let us recall that the starting point for all our analysis has been the differential relation, \(d\psi_{ness} = X_i d\lambda^i\) which allowed us to identify the physical variables as being local coordinates on a contact manifold. However, this identification is not unique in the sense that one can always multiply eqn (\ref{firstlaw}) with a non-zero function \(f\) on the thermodynamic phase space and still perform a similar analysis, i.e. take
\begin{equation}\label{genfl}
  f(d\psi_{ness} - X_i d\lambda^i) = 0
\end{equation} on \(Q\) as the starting point. For example, if \(f = 1/X_1\), one would get
\begin{equation}\label{firstlaw2}
  -d\lambda^1 + \frac{d\psi_{ness}}{X_1} - \frac{X_j}{X_1}d\lambda^j = 0
\end{equation} where \(j = 1,2,....,n-1\). Comparison of the above relation with eqn (\ref{contactvanish}) leads us to the identification that \((\psi_{ness},\lambda^j)\) are independent coordinates on the control parameter space with generating function (with a minus sign) \(\lambda^1=\lambda^1(\psi_{ness},\lambda^j)\) obtained by inverting \(\psi_{ness}=\psi_{ness}(\lambda^1,\lambda^j)\) for \(\lambda^1\). Note that eqns (\ref{firstlaw}) and (\ref{firstlaw2}) are physically equivalent. However, following eqn (\ref{metriclocal}), the new thermodynamic length shall be defined as,
\begin{equation}\label{hessianX1}
  dl_1^2 = -\frac{\partial^2 \lambda^1}{\partial \chi^i \partial \chi^j} d\chi^i d\chi^j, \hspace{3mm} i,j \in \{1,2,....,n\}
\end{equation} with \(\chi^1 = \psi_{ness}\). It is then a natural question to ask whether or not the thermodynamic lengths described by the metrics given in eqns (\ref{metric}) and (\ref{hessianX1}) are related. In this context, we state the following result,
\begin{theorem}\label{theoremconformal}
  The thermodynamic lengths are conformally related as,
  \begin{equation}\label{conformal}
    dl^2 = X_1 dl_1^2.
  \end{equation}
\end{theorem}
Its proof can be understood as follows. Recall that on a generic Legendre submanifold \(L\), the metric is given by \(G_L = dq^i dp_i\). Starting with eqn (\ref{firstlaw}), for independent parameters \(\{\lambda^i\}\) with generating function \(\psi_{ness}\) this reads,
\begin{equation}
  dl^2 = dX_i d\lambda^i
\end{equation} whereas, if eqn (\ref{firstlaw2}) is taken as the starting point, then the independent parameters are \((\psi_{ness},\lambda^j)\) with dependent parameters \((-1/X_1, X_j/X_1)\) leading to the line element,
\begin{equation}
  dl_1^2 = d(-1/X_1) d\psi_{ness} + d(X_j/X_1) d\lambda^j.
\end{equation} With a slight rearrangement and substituting eqn (\ref{firstlaw2}), it follows that the last relation simplifies to,
\begin{equation}
  dl_1^2 = \frac{1}{X_1} dX_i d\lambda^i
\end{equation} thus proving the relationship between \(dl^2\) and \(dl_1^2\) given in eqn (\ref{conformal}).

\smallskip

We emphasize on the fact that we haven't used any particular form of the control variables or response functions for deriving this result. In fact, if we take \(f = 1/X_2\) in eqn (\ref{genfl}) and label the resulting thermodynamic length \(dl_2^2\) and similarly take \(f = 1/X_k\) for some \(k \in \{1,2,....,n\}\) and then label the corresponding thermodynamic length to be \(dl_k^2\), the following result can be straightforwardly shown as a consequence of theorem-(\ref{theoremconformal}),
\begin{corollary}
  The line element \(dl^2\) is conformally related to the entire set of line elements \(\{dl_i^2\}\) as,
  \begin{equation}\label{gencon}
    dl^2 = X_1 dl_1^2 = X_2 dl_2^2 = .... = X_k dl_k^2 = .... = X_n dl_n^2.
  \end{equation}
\end{corollary}
Since, \(dl^2\) is the length associated with the Schl\"{o}gl metric and is equivalent to the Fisher information matrix, thus for a general system, eqns (\ref{gencon}) may provide alternate means of computing the thermodynamic length consistent with Fisher's information matrix.
%\subsection{Relationship with friction tensor}

%\subsection{Finite time transformations}

\section{Discussion}
Investigation of the geometry of the non-equilibrium thermodynamic phase space, based upon the differential geometric approach provides a deeper understanding of the structure of thermodynamics and statistical mechanics. For steady state quantum systems arbitrarily far from equilibrium, we have established that NESS are points on control parameter spaces which are Legendre submanifolds of an ambient non-equilibrium thermodynamic phase space with an underlying contact structure. The structure presented in this work crucially depends on the introduction of the steady-state Massieu--Planck function, identification of its thermodynamic relation (and non-equilibrium Maxwell/Onsager relations) as well as finding the correct set of control parameters in~\cite{taniguchi}.
The machinery of contact geometry and Hamiltonian dynamics can then be successfully applied to NESS and reveals a rich geometric structure as is contained in section-(III) of the present paper. Henceforth, our results are encouraging in the sense, it may help us to understand the non-equilibrium thermodynamic steady states and their geometric structure. One can also expect to understand in a better way the optimal protocol and efficiency of biological systems, synthetic molecular machines and kinetic phenomena like (bio)chemical reactions from the point of view of contact Hamiltonian dynamics.

\smallskip

In the context of non-equilibrium thermodynamics and in particular where the processes are the steady state ones, the physical meaning of the thermodynamic metrics and their curvature scalars deserve to be studied in detail. For systems at thermodynamic equilibrium, it is an established fact that the curvature scalar associated with the Ruppeiner metric is positive or negative respectively (in a sign convention opposite to that adapted in \cite{Janyszek:1989zz}) for the cases where the microscopic degrees of freedom interact in a repulsive or attractive manner \cite{interactions}. Between these two regimes is the ideal gas limit, where the system is non-interacting and the curvature scalar comes out to be zero identically. It is then worthwhile to study the physics captured by the thermodynamic curvature in case of systems driven out of equilibrium. In particular, since it is the control parameters that one manipulates externally, of outmost importance is to study the physical implications of the scalar curvature associated with the metric \(g_{ij}\) [eqn (\ref{metric})] defined on control parameter spaces. These issues should be addressed in the future.
%% The Appendices part is started with the command \appendix;
%% appendix sections are then done as normal sections
\

\section*{Acknowledgements}
We are grateful to Rishabh Raturi for valuable discussions. A.G. would like to acknowledge the support by the International Centre for Theoretical Sciences (ICTS) for the online program - Bangalore School on Statistical Physics - XI (code: ICTS/bssp2020/06) and the financial support received from the M.H.R.D., Government of India in the form of a Prime Minister's Research Fellowship. M.B. gratefully acknowledges financial support from Department of Science and Technology (DST), India under
the Core grant (Project No. CRG/2020//001768). C.B. gratefully acknowledges the support received from S.E.R.B., Government of India, MATRICS (Mathematical Research Impact Centric Support) grant no. MTR/2020/000135. The authors are grateful to the anonymous referee for valuable comments which have led to a substantial improvement of article.

%% If you have bibdatabase file and want bibtex to generate the
%% bibitems, please use
%%
% \bibliographystyle{elsarticle-num}
% \bibliography{cas-refs}

%% else use the following coding to input the bibitems directly in the
%% TeX file.

\end{document}